\documentclass{moriond}

\def\be{\begin{equation}}
\def\ee{\end{equation}}
\def\bea{\begin{eqnarray}}
\def\eea{\end{eqnarray}}

\begin{document}
\newcommand{\Photo}{}
\vspace*{4cm}
\title{GRAVITATIONAL BIREFRINGENCE OF LIGHT AT COSMOLOGICAL SCALES}

\author{ CHRISTIAN DUVAL and THOMAS SCH\"UCKER }

\address{Aix Marseille Univ, Universit\'e de Toulon, CNRS, CPT, Marseille, France}

\maketitle\abstracts{
Birefringence of light induced not by matter, but by the gradient of an electric field, has been predicted in 1955 and observed in 2008. Here we replace the electric field by the gravitional field of our expanding universe.}

\section{The Fedorov-Imbert effect}

Considering the boundary between air (or vacuum) and a glass plate as a discontinuous electric field, Fedorov~\cite{Fed55} and Imbert~\cite{Imb72} predicted that photons of a given polarisation follow a discontinuous trajectory with an offset of the order of a wave length as shown in Figure 1. Note that the infinite gradient of the electric field induces an infinite speed of light along the dotted line. The offset of photons of the opposite polarisation is in the opposite direction. Therefore photons of different polarisations follow different trajectories, an effect called birefringence or also Spin Hall Effect of Light.

\begin{figure}[h]
\centerline{\includegraphics[scale=.7]{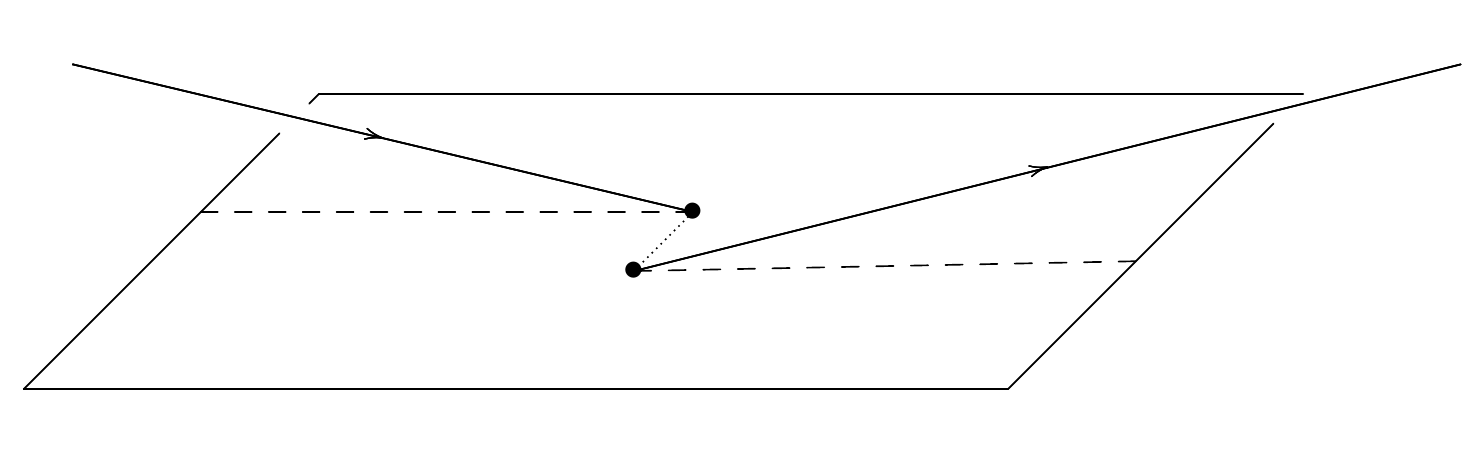}}
\caption[]{The Fedorov (1955) Imbert (1972)  effect for reflection: A plane glass surface reflects an incoming, circularly  polarized  light beam. The dashed lines indicate the orthogonal projections of incoming and reflected light beams onto the glass surface. The dotted line (between the blobs) is the offset between incoming and reflected beams. It is of the order of the wavelength of the light beam.}
\label{FI}
\end{figure}
This effect was observed for the first time 10 years ago~\cite{BNKH08,HK08} using techniques of weak quantum measurement.

\section{Adding spin to geodesics}

Gravitational fields with gradient also induce birefringence of light. To see this we must generalize the geodesics, which describe trajectories of point-like test particles without spin in a gravitational field.  The gravitational field is encoded in the Christoffel symbols of a pseudo-Riemannian metric. 

Let $X^\mu (\tau)$ be the trajectory of a massless spinless particle with 
4-velocity 
$dX^\mu/d\tau$ and
4-momentum
$P^\mu(\tau) $.
Its equations of motion (geodesics) read in first order formalism, 
\begin{eqnarray}
\frac{d}{d\tau}\,X^\mu  &=&P^\mu,
\\[1mm]
\frac{D}{d\tau}\,P^\mu &=&0,\end{eqnarray}
with $D/d\tau$ denoting the covariant derivative with respect to the Christoffel symbols. The equations of motion 
have one conserved quantity, $P^\mu P_\mu =m^2=0$. Note that the massless limit is delicate; by the equivalence principle the geodesics for massive particles do not depend on this positive mass $m$ and the limit $m\rightarrow 0$ cannot be continuous.

In the seventies spin was added to the massless particle~\cite{Kun72,DFS72,Sou74,Sat76} in the form of the 
\textsl{antisymmetric} {spin tensor} $S^{\mu\nu}(\tau)$ coupled to the gradient of the gravitational field encoded in the Riemann tensor.
\begin{eqnarray}
\frac{d}{d\tau}\,X^\mu  &=&P^\mu +2\,\frac{{S^\mu }_\nu {R^\nu}_{\beta \rho \sigma }S^{\rho \sigma }P^\beta }{R_{\alpha \beta \rho \sigma }S^{\alpha \beta }S^{\rho \sigma }}\,, \label{mot1}
\\[1mm]
\frac{D}{d\tau}\,P^\mu &=&-s\,\frac{\sqrt{-\det({R^\alpha }_{\beta \rho \sigma }S^{\rho \sigma })}}{R_{\alpha \beta \rho \sigma }S^{\alpha \beta }S^{\rho \sigma }}\,P^\mu \,,
\\[4pt]
\frac{D}{d\tau}\,S^{\mu \nu}&=&P^\mu \frac{d}{d\tau}\,X^\nu-P^\nu \frac{d}{d\tau}\,X^\mu\,,\label{mot3}
\end{eqnarray}
where $s$ is the {\it ``scalar spin''} defined by
\begin{equation} 
 -{\textstyle\frac{1}{2}} {S^\mu }_\nu{S^\nu }_\mu=: s^2\,.
 \end{equation}
 Let us anticipate that upon a 3+1 split $X=(\vec x,t)$, the scalar spin becomes the projection of the spin 3-vector $\vec s$ onto the 3-moment $\vec p$ (or the highest and lowest weight of the spin representation). Photons have $s=\pm\hbar.$

In presence of spin, we have three more conserved quantities, the scalar spin $s$ and
  \begin{equation} 
   P_\mu\,\frac{d}{d\tau}X^\mu  =0,\quad
    {S^\mu }_\nu P^\nu=0.
  \end{equation}  
  The equations of motion (\ref{mot1} - \ref{mot3})  for the massless particle with spin have three delicate properties:
  \begin{itemize}\item
  They are ill defined in Minkowski space because of the vanishing Riemann tensor.
  \item
  The limit of vanishing spin is delicate as for the limit of vanishing mass in the case of geodesics.
  \item
  We are confronted with super-luminal propagation velocities as in the Fedorov-Imbert effect.
  \end{itemize}
    Note also that all four conserved quantities are valid for an arbitrary metric, they do not  derive from isometries via Emmy Noether's theorem. A self-contained derivation of the equations of motion from the massive Mathisson-Papapetrou-Dixon equations can be found in the appendix of reference~\cite{k0}.
  
  \section{The example of the Robertson-Walker metric}
  
  In 1976 Saturnini~\cite{Sat76} has analyzed the equations of motion (\ref{mot1} - \ref{mot3}) for the Schwarzschild metric. Here we consider flat Robertson-Walker metrics~\cite{k0}. 
  
  Because of the many vector products in the  computations we use
 Euclidean coordinates $\vec x$ and   cosmic time $t$ and write the line element as
$
-a(t)^2\,d\vec x^2+dt^2,
$
with positive scale factor $a>0$, that we also suppose increasing, $a'>0$. Thanks to Emmy Noether's theorem for the six Killing vectors for translations $\partial/\partial\vec x$ and rotations $\vec x\,\wedge\,\partial/\partial\vec x$ and the conformal Killing vector $a(t)\,\partial/\partial t$, the equations of motion reduce drastically and can be solved by a Runge-Kutta algorithm.
Figure 2 shows a typical result in the $\Lambda CDM$ model. For a given polarisation, $s=\hbar$, the trajectory of the photon is the helix. The dashed line is the trajectory of a `photon' without spin, $(x^1(t),0,0)$. The transverse spin $\vec s^\perp_e$ at emission time $t_e$ is indicated by the short arrow at the left. The opposite polarisation, $s=-\hbar$, produces the same helix but of opposite chirality. The spin $\vec s$ precesses around the center of the helix with the same variable instantaneous period as the period of the helix. 
 \begin{figure}[h]
\begin{center}
\includegraphics[scale=0.54]{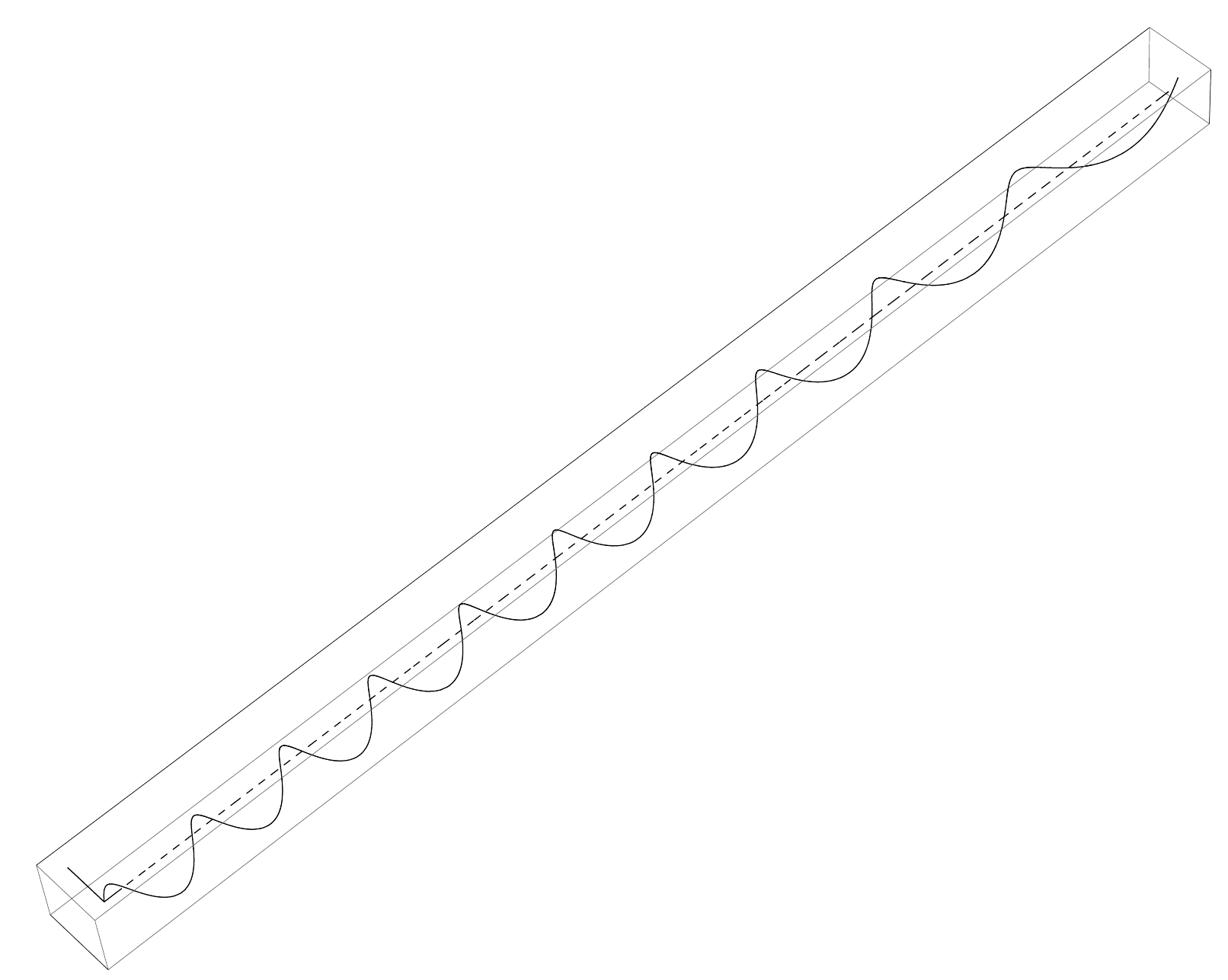}
\caption {The trajectory of photons, $\vec x(t)$, in a flat Robertson-Walker universe in comoving coordinates is the helix. The dashed line is the null geodesic. 
The transverse spin $\vec s^\perp_e$ at emission time $t_e$ is indicated by the short arrow at the left.
\label{helix}
}
\end{center}
\end{figure}
Let us denote this period by $T_\mathrm{helix}(t)$. We also want to compute the position of the center $\vec x_\mathrm{center}(t)$ of the helix and its radius $R_\mathrm{helix}(t)$.  To this end we linearize the equations of motion 
 in the parameter $|\vec s_e|\,\lambda _e/(2\pi \,\hbar\,a_e)$, $\lambda _e$ being the wavelength at emission. For a Lyman $\alpha $ photon of redshift 2.4 this parameter is of the order of $10^{-34}$, well justifying the linear approximation. It yields:
\begin{equation}
T_\mathrm{helix}(t)
\sim\frac{a(t)}{a_e}\,\frac{\lambda _e}{1+q(t)}\quad {\rm with}\quad q:=-a\,a''(t)/a'(t)^2.
\end{equation}
\begin{equation}
\vec x_\mathrm{center}(t)
\sim
\left(
\begin{array}{c}
 x^1(t)\\
0\\
-\,\frac{\lambda _e}{2\pi\,a_e}\,\left(1-a'_e{}x^1(t)\right)
\end{array}\right),\quad
R_\mathrm{helix}(t)
\sim\frac{a(t)}{a_e}\,\frac{\lambda _e}{2\pi}.
\end{equation}
Note that in this approximation the projection of the helix on the null geodesic coincides with this geodesic at all times. 
To be concrete, consider a $z=2.4$ Lyman $\alpha $ photon engaged in a race with a fictitious mass- and spinless competitor travelling at the speed of light. Both competitors are emitted by the same source at the same time, the race lasts $3\cdot 10^9$ light years and they arrive simultaneously in our telescope.  However, they arrive with a transverse offset of the order of $10^{-7}$ meter. To achieve this remarkable draw, the photon on its helix has to travel at roughly $\sqrt{2} $ times the speed of light during the entire race. 

Said differently, the superluminal propagation in the equations of motion (\ref{mot1} - \ref{mot3}) does violate causality, but only at the tiny scale of the order of a wave length. For a quantum of solace, note that the support of the Feynman propagator of the Dirac operator in Minkowski space leaks out of the light-cone~\cite{scharf}. However this leakage is damped exponentially and also there, causality violation is extremely tiny.

Robertson-Walker metrics with curvature give similar results as computed in reference~\cite{dpst}. There you also find an unsuccessful attempt on finding spin effects in the Hubble diagram of supernovae.

\section{Conclusions and questions}

\begin{itemize}\item
 The gravitational field of an expanding universe produces birefringence of light.
\item
This birefringence carries information on the acceleration of the universe.
\item
Of course the main question is: Can this birefringence be measured?
  \end{itemize}
In 1976 Saturnini~\cite{Sat76} has shown that the gravitational field of a static, spherical mass produces birefringence of light.
  \begin{itemize}\item
How does birefringence in the Schwarzschild solution interfere with lensing?
\end{itemize}
Einstein predicted gravitational waves in 1916. They were observed in 2015. 
Concerning birefringence, three questions are immediate:
  \begin{itemize}\item
Does the gravitational field of a gravitational wave also produce birefringence of light?
\item
If yes, what information is carried by this birefringence?
\item
Can this birefringence be measured in interferometers with polarized laser beams?
\end{itemize}
Taking due account of its spin, the photon propagates through an expanding universe on a helix. Did Feynman already know about this when he chose to represent the photon propagator by a curly line?

\section*{Acknowledgments}

As always, it is a pleasure to thank Vera de S\'a-Varanda, Jacques Dumarchez and their team for their warm and efficient hospitality.

\end{document}